\documentclass[aps,prb,floatfix]{revtex4}

\usepackage[]{graphicx}

\begin{document}


\title[Photoluminescence of nanocrystals embedded in oxide matrices]{Photoluminescence of nanocrystals embedded in oxide matrices}

\author{C. Estrada, J.A. Gonzalez, A. Kunold}
\affiliation{Ciencias B\'asicas, UAM-Azcapotzalco, Av. S. Pablo 180, C. P. 02200, M\'exico D. F., M\'exico.\\ email akb@correo.azc.uam.mx}
\author{J. A Reyes-Esqueda}
\affiliation{Instituto de F\'isica, UNAM, M\'exico D. F., M\'exico}
\author{P.Pereyra}
\affiliation{Ciencias B\'asicas, UAM-Azcapotzalco, Av. S. Pablo 180, C. P. 02200, M\'exico D. F., M\'exico.}
\begin{abstract}
We used the theory of finite periodic systems to explain the
photoluminescence spectra dependence on the average diameter of
nanocrystals embedded in oxide matrices. Because of the broad matrix
band gap, the photoluminescence response is basically determined by
isolated nanocrystals and sequences of a few of them. With this
model we were able to reproduce the shape and displacement of the
experimentally observed photoluminescence spectra.

\end{abstract}

\maketitle

\section{Introduction}

ZnO-Ge and SiO2-Ge nanocomposites are intensively studied for their
possible use as optoelectronic devices
\cite{Fan,Jin,Bagnall,Takeoka,Chu}. A common method for preparing
these systems is by alternate deposition of oxides and Ge, followed
by an annealing process to increase the diffusion of nanocrystalites
(NC) into the adjacent
layers\cite{Takeoka,Maeda1,Kanemitsu,Maeda2,Okamoto}. This process
leads also to homogenize the sizes and spacial dispersion of NCs. It
is widely known that the photoluminescense-spectra shift is strongly
related to NC's size. Although there have been several attempts to
understand the relation between the morphology and the
photoluminescence (PL) response of embedded NC in oxide matrices
\cite{Fan,Takeoka,Maeda1}, none of them lead to a simple and precise
explanation. The purpose of this work is to understand and to
explain this relation from a simple model and a few physical
parameters.

To calculate the optical response of these rather complex systems,
we use both the {\it bona fide} multiple-quantum-well electron and
hole eigenvalues and eigenfunctions, obtained from the theory of
finite periodical systems\cite{Pereyra} (TFPS) and a simple
assumption, experimentally supported, on the statistical
distributions of NC sizes. We show that the main features of the PL
spectra (in Ref. \cite{Fan} for $Ge$ doped $ZnO$ films annealed at
several temperatures), can clearly be identified in terms of the
interband excitonic transitions calculated for multiple quantum well
(MQW) systems, whose well and barrier parameters are drawn from the
experimental system. The agreement with other kind of experimental
results, as those reported by by S. Takeoka et al. \cite{Takeoka}
for $Ge$ NC in $SiO_2$, 
is also good
when the nanocrystals-size dispersion is taken into account.

\section{The model}
\begin{figure}
\includegraphics[width=2.0in]{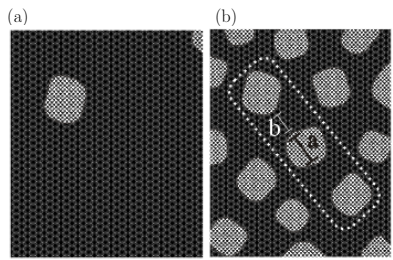}
\caption{Ge cristalites embedded in a ZnO matrix for (a) low NCs densities and
(b) high densities. } \label{fig1}
\end{figure}

\begin{figure}
\includegraphics[width=4.5in]{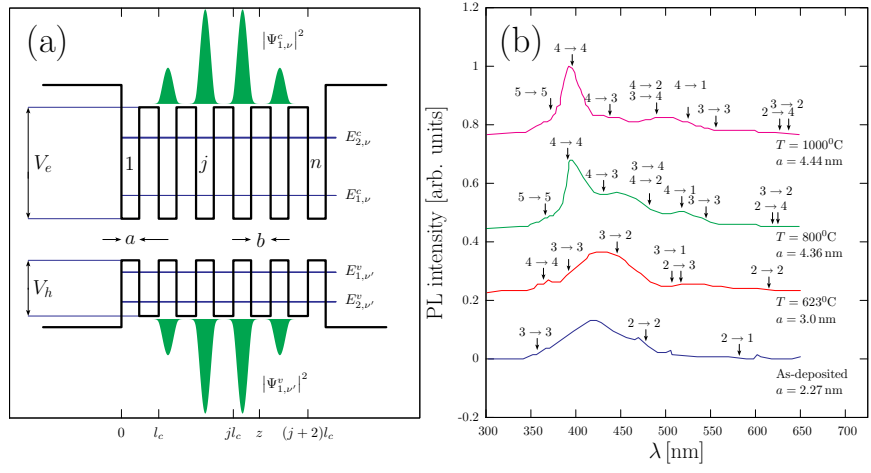}
\caption{
(a) Finite periodic square-well modulation of the valence and
conduction band edges along a set of $n$ QW's. Also we show the wave
functions $\Psi^c_{1,1}$ and $\Psi^v_{1,1}$ for the first energy
level of the first subband in the conduction and valence band. For
these wave functions the electron and hole are localized inside the
QW's.
(b) Room temperature PL spectra of Ge doped ZnO films annealed
at several temperatures and theoretical peaks position (black
arrows). We also show the corresponding NC's average diameter used
for the theoretical model.} \label{fig2}
\end{figure}

In Fig. \ref{fig1} we observe micrographs of Ge cristalites
embedded in a ZnO matrix. At low NC densities we have almost
isolated NCs (see Fig \ref{fig1} (a)), while at higher NC
densities one can find groups of NCs aligned and almost
equidistant(see Fig. \ref{fig1} (b)). Although a single NC
can be better modeled as a quantum dot, the relevance of the
longitudinal coherence in the energy spectra of a sequence of NCs,
leads us to model the sequence as a multiple 1D-quantum well
potential. Each of these sequences is then treated as a finite
periodic square-well potential (see Fig. \ref{fig2} (a)), and the
whole nanocomposite PL response results from the contribution of an
ensemble of MQWs, which differ basically in the number of NCs,
well-widths and barrier-thicknesses.

In Fig. \ref{fig2} (a) we have a MQW potential for a given set of
parameters. We show the energy spectra (in the conduction and
valence bands), and the square of the wave function amplitude
$\psi_{1,1}^{c(v)}$ for the first energy level in the first conduction
(valence)
subband. In the TFPS, these quantities can easily be calculated. The
eigenvalues are obtained from \cite{Pereyra}
\begin{equation}
(\alpha_{n}e^{ika}-\alpha_{n}^{*}e^{-ika})+\beta_{n}-\beta_{n}^{*}=0,
\label{e.1}
\end{equation}
with $k=\sqrt{2mE/\hbar^2}$ and $m$ the electron or hole effective
mass. The wave functions at any point $z$ in the $(j+1)$-th cell
(for a system of length $L=nl_c$) and energy $E$ are given by
\begin{equation}
\Psi^{\iota}(z,E)=A[(\alpha_{p}+\gamma_{p})(\alpha_{j}-\beta_{j}\frac{\alpha_{n}+\beta_{n}^{*}}{\alpha_{n}^{*}+\beta_{n}})
+(\beta_{p}+\delta_{p})(\beta_{j}^{*}-\alpha_{j}\frac{\alpha_{n}+\beta_{n}^{*}}{\alpha_{n}^{*}+\beta_{n}})].
\label{e.2}
\end{equation}
 $\alpha_{j},\beta_{j}$ are elements of the transfer matrix
$M(jl_c,0;E)$ and $\alpha_{p}, \beta_{p}, \gamma_{p}, \delta_{p}$
are elements of a partial cell transfer matrix $M_{p}(z,jl_c;E)$.
\begin{figure}
\includegraphics[width=4.8in]{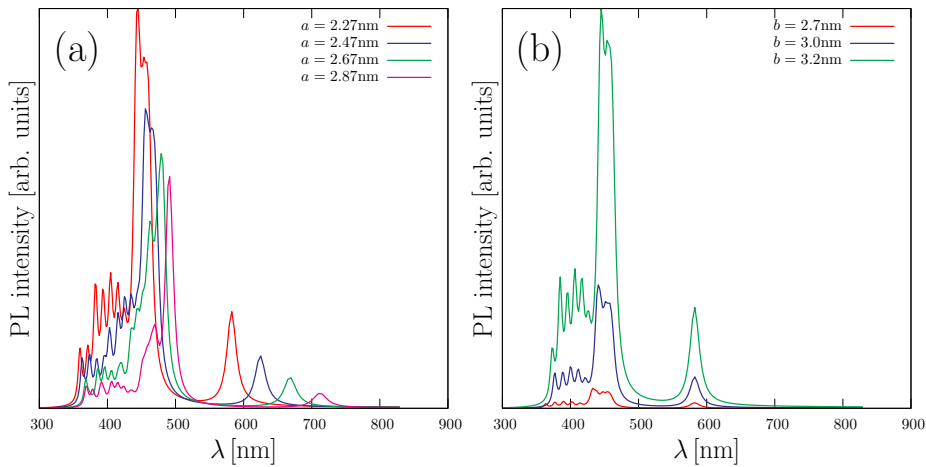}
\caption{Theoretical PL spectra of Ge doped ZnO films for (a) several
NCs diameters and (b) barrier widths.} \label{fig3}
\end{figure}
\begin{figure}
\includegraphics[width=4.9in]{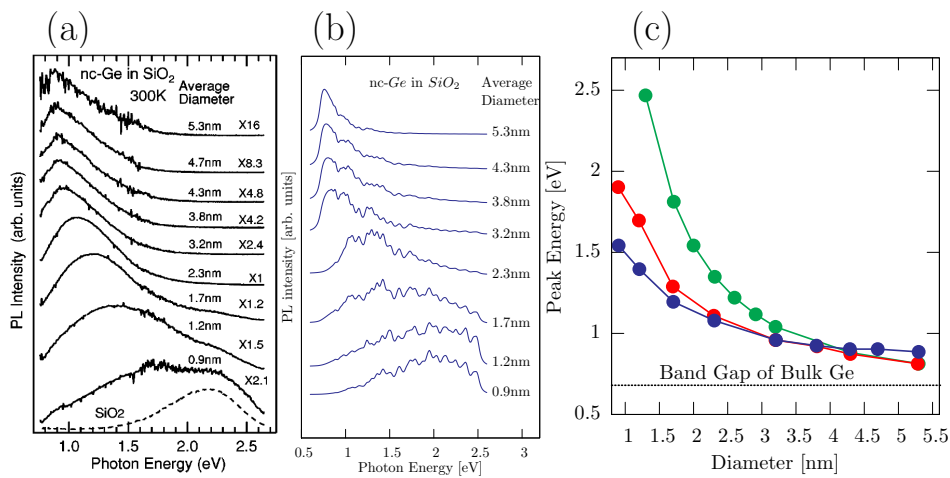}
\caption{At the left the measured PL response for different average
diameters of NC-Ge embedded in $SiO_2$. In the middle our
theoretical curves, and at the right image we plot the PL peak
energy versus average diameter of NC-Ge for:  experiment
\cite{Takeoka},  the theoretical prediction with
size-dispersion and  without size-dispersion.} \label{fig4}
\end{figure}
$\Psi^{\iota}(z,E_{\mu,\nu}^{\iota})$ (with $\iota=c,v$ for
conduction and valence bands), is the eigenfunction corresponding to
the eigenvalue $E_{\mu,\nu}^{\iota}$. To evaluate the PL response we
use the following function
\begin{equation}\label{susceptibilidad}
\chi=\sum_{\nu,\nu',\mu,\mu'} \frac{p_{\nu,\nu',\mu,\mu'}
f\left(E^c_{\mu,\nu}\right)
\left[1-f\left(E^v_{\mu\prime,\nu\prime}\right)\right]}
{(w_{s}-E_{\mu\nu}^c+E_{\mu'\nu'}^v)^{2}+\Gamma^{2}}.
\end{equation}
 $f(E^\iota_{\mu,\nu})$ the Fermi-Dirac
carrier distribution functions in the conduction or valence band.

\section{Results and discussion.}
In Fig. \ref{fig2} (b) we present the experimental PL spectra of
$Ge$-doped $ZnO$ films for different annealing temperatures,
reported in Ref. \cite{Fan}. The NC's average diameter increases
with the annealing temperature \cite{Maeda2}. The peaks position in
these PL curves can be described in terms of the energy spectra of a
single sequence of NCs. For this calculation we use the potential
and gap parameters corresponding to the $\Gamma$ point of $ZnO$ and
the annealing temperature, i.e. $m^e_{Ge}=0.12 m_e$, $m^h_{Ge}=0.23
m_e$, $m^e_{ZnO}=0.24 m_e$, $m^h_{ZnO}=0.45 m_e$, $E_{Ge}=0.66 {\rm
eV}$,$E_{ZnO}=3.31{\rm eV}$. The black arrows point out the
theoretical predictions for different possible recombination
transitions. In this figure, the assignment $\mu \rightarrow \mu$
refers to the energy transitions $E_{\mu,\nu}^c \rightarrow
E_{\mu,\nu}^v$. It is well known that increasing the well width $a$,
the subbands move towards lower energies while their number
increases. We notice in the the experimental PL curves that
increasing $a$ the small peaks shift toward larger wave lengths,
while the main PL peak displaces in the opposite direction. This
behavior is even much clear in the experimental results of Takeoka
eta al. in Ref. \cite{Takeoka}, where the secondary-peaks structure
is absent.

In these systems the experimental curves arise from the contribution
of a large number of NCs which sizes are statistically
distributed\cite{Maeda2}. The PL aspect and peaks relative-height
depend not only on the potential parameters, but also on their
statical distribution. The PL response of a single MQW in Fig.
\ref{fig3}, shows the red-shift with $a$ and the high PL sensitivity
to small barrier-thickness variations.

To explain the experimental results reported by Takeoka et al. in
Ref \cite{Takeoka} for $Ge$ Ncs in a $Si0_2$ matrix, we need to take
into account also the NC's size distribution. We assume, according
with the experimental results, that the size distribution is
described by a log-normal function. We then calculate the PL spectra
of several sets of NC diameters, pondered by their statistical
weight. Since the PL peaks position is less sensitive to  the number
of wells and to the NC spacing, we assume for this case a fixed
barrier width and also a fixed number of cells.

At Fig. \ref{fig4} (a) we present the experimental results
from Ref. \cite{Takeoka}. In Fig. \ref{fig4} (b) we show
our calculated curves assuming up to $50$ contributing MQW sequences
with the following parameters: $m^e_{SiO_2}=0.5 m_e$, $m^h_{SiO_2}=
0.49 m_e$,$E_{SiO_2}=9.3 {\rm eV}$, corresponding also to the
$\Gamma$ point of $SiO_2$. At Fig. \ref{fig4} (c), we plot
also the peak position versus the NCs average diameter. This
averaging procedure improves the agreement with the experimental PL
spectra. The size-dispersion shifts the PL peaks toward lower
energies. This is a consequence of the asymmetry of the log-normal
distribution.

In our calculation we neglected {\it e-h} Coulomb interaction
corrections. In any case, they will produce small red shifts in the
secondary PL peak positions and will be washed out in the main peak
behavior.

\section{Conclusions}

To explain the photoluminescence response of NCs embedded in oxide
matrices as function of the average NCs diameter, we modeled these
systems as a collection of 1-D multiple-quantum-well potentials.
Using the finite periodical systems theory and a simple NC-size
distributions assumption, we shown that the calculated 1-D MQW
subband transitions, describe correctly the secondary PL peaks
behavior, observed by Fan et al. \cite{Fan}, as function of the NCs
diameter. We have shown also that the main PL peak blue-shift
behavior, observed by Takeoka et al. in Ref. \cite{Takeoka}, is
basically determined by the statistical dispersion of the NCs
diameter.

\end{document}